\documentclass[11pt]{article}
\usepackage{amsmath,amssymb,amsthm,color,graphicx}
\usepackage{subfigure}

\newtheorem{theorem}{Theorem}
\newtheorem{lemma}{Lemma}
\newtheorem{claim}{Claim}

\newtheorem{corollary}{Corollary}

\theoremstyle{definition}

\newcommand{\bbox}{\vrule height7pt width4pt depth1pt}

\newcommand{\comment}[1]{}
\newcommand{\ignore}[1]{}

\def\clap#1{\hbox to 0pt{\hss#1\hss}}

\makeatletter
  \def\moverlay{\mathpalette\mov@rlay}
  \def\mov@rlay#1#2{\leavevmode\vtop{%
    \baselineskip\z@skip \lineskiplimit-\maxdimen
    \ialign{\hfil$#1##$\hfil\cr#2\crcr}}}
\makeatother

\def\P{{\cal P}}

\newcommand{\wdg}[1]{wedge({#1})}

\newcommand\CH[1]{\ensuremath{\mathcal{CH}\left(#1\right)}}

\newcommand{\remove}[1]{}

\begin{document}
\title{Ice-Creams and Wedge Graphs}
\author{
Eyal Ackerman
\and
Tsachik Gelander
\and
Rom Pinchasi
}

\date{}

\maketitle

\sloppy

\begin{abstract}
What is the minimum angle $\alpha >0$ such that given any set of 
$\alpha$-directional antennas (that is, antennas each of which 
can communicate along a wedge of angle $\alpha$), one can always assign a 
direction
to each antenna such that the resulting communication graph is connected?
Here two antennas are connected by an edge if and only if each lies in the
wedge assigned to the other.
This problem was recently presented by Carmi, Katz, Lotker, and Ros\'en~\cite{CKLR10} 
who also found the minimum such $\alpha$ namely 
$\alpha=\frac{\pi}{3}$. In this paper we give a simple %strikingly simple and elegant
proof of this result. Moreover, we obtain a much stronger and optimal result (see Theorem~\ref{theorem:main}) saying in particular that one can chose
the directions of the antennas so that the communication graph has diameter 
$\le 4$.

Our main tool is a surprisingly basic geometric lemma that is of 
independent interest.
We show that for every compact convex set $S$ in the plane and every $0 < \alpha < \pi$, 
there exist a point $O$ and two supporting lines to $S$ passing 
through $O$ and touching $S$ at two \emph{single points} 
$X$ and $Y$, respectively,
such that $|OX|=|OY|$ and the angle between the two lines is $\alpha$.
\end{abstract}

\section{Antennas, Wedges, and Ice-Creams}

Imagine the following situation. You are a manufacturer of antennas.
In order to save power, your antennas should communicate along a 
wedge-shape area, that is, an angular and practically infinite 
section of certain angle $\alpha$ 
whose apex
is the antenna. The smaller the angle is the better it is in terms of power
saving. You are supposed to build many copies of these antennas to be used
in various different communication networks. You know nothing about the future 
positioning of the antennas and you want them to be generic in the sense
that they will fit to any possible finite set of locations.
When installed, each antenna may be directed to an arbitrary direction that
\begin{figure}[h]
    \centering
    \includegraphics[width=7cm]{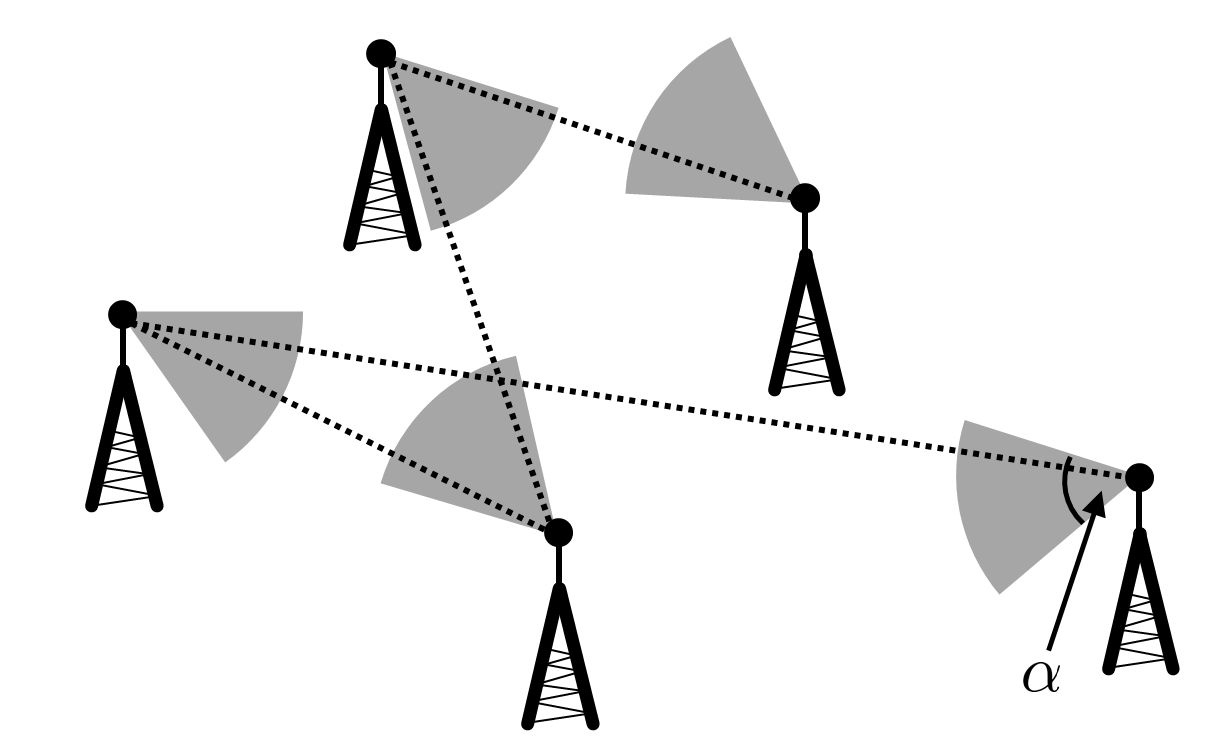}
	\caption{a set of $\alpha$-directional antennas and their (connected) communication graph.}
	\label{fig:antennas}
\end{figure}
will stay fixed forever. Therefore, you wish to find the minimum $\alpha > 0$
so that no matter what finite set $P$ of locations of the antennas is given, 
one can always install the antennas and direct them so that they can communicate with 
each other. This is to say that the \emph{communication graph} of the antennas 
should 
be a connected graph. The communication graph is the graph whose vertex set
is the set $P$ and two vertices (antennas) are connected by an edge
if the corresponding two antennas can directly communicate with each other,
that is, each is within the transmission-reception wedge of the other
(see Figure~\ref{fig:antennas} for an illustration).

This problem was formulated by Carmi, Katz, Lotker, and Ros\'en~\cite{CKLR10}, 
who also found the optimal $\alpha$ which is 
$\alpha=\frac{\pi}{3}$ (a different model of a \emph{directed} communication 
graph of directional antennas
of bounded transmission range was studied in~\cite{CKKKW08,DF10,DKKOPS10}).  

In this paper we provide a much simpler and more elegant proof of this result.
We also improve on the result in~\cite{CKLR10} by obtaining a much simpler 
connected communication graph for $\alpha=\frac{\pi}{3}$ whose diameter
is at most $4$. Our graph in fact consists of a path of length $2$
while every other vertex is connected by an edge to one of the three
vertices of the path.

In order to state our result and bring the proof we now formalize some of the
notions above.

Given two rays $q$ and $r$ with a common apex in the plane, we denote by $\wdg{q,r}$ 
the closed convex part of the plane bounded by $q$ and $r$.
For three noncollinear points in the plane $A,B,C$ we denote by
$\angle ABC$ the wedge $\wdg{\overrightarrow{BA}, \overrightarrow{BC}}$,
whose apex is the point $B$. $\angle ABC$ is a wedge of angle
$\measuredangle ABC$.

Let $W_{1}, \ldots, W_{n}$ be $n$ wedges with pairwise distinct apexes.
The \emph{wedge-graph} of $W_{1}, \ldots, W_{n}$ is by definition the 
graph whose vertices correspond to the apexes $p_{1}, \ldots, p_{n}$
of $W_{1}, \ldots , W_{n}$, respectively, where two apexes $p_{i}$ and $p_{j}$ are joined by an 
edge iff $p_{i}  \in W_{j}$ and $p_{j} \in W_{i}$.

Using this terminology we wish to prove the following theorem whose first part
was proved by Carmi {\em et al.}~\cite{CKLR10}.

\begin{theorem}\label{theorem:main}
Let $P$ be a set of $n$ points in general position in the plane
and let $h$ be the number of vertices of the convex hull of $P$.
One can always find in $O(n \log h)$-time $n$ wedges of angle $\frac{\pi}{3}$ whose 
apexes are the $n$ points of $P$ such that the wedge-graph with respect 
to these wedges is connected. Moreover, we can find wedges so that the 
wedge graph consists of a path of length $2$ and each of the other vertices
in the graph is connected by an edge to one of the three vertices of the path.
\end{theorem}

The angle $\frac{\pi}{3}$ in Theorem~\ref{theorem:main} is best possible,
as shown in \cite{CKLR10}.
Indeed, for any $\alpha < \frac{\pi}{3}$
one cannot create a connected communication graph for a set of $\alpha$-directional antennas
that are located at the vertices of an equilateral triangle and on one of its edges.

We note that the result in Theorem \ref{theorem:main} is optimal in the sense
that it is not always possible to find an assignment of wedges to the points
so that the wedge graph consists of less than three vertices the union of 
neighbors of which is the entire set of vertices of the graph.
To see this consider a set of points evenly distributed on a circle.
Notice that if each wedge is of angle $\alpha \leq \frac{\pi}{3}$,
then in any wedge graph each vertex is a neighbor of at most one third of 
the vertices.

% Enough for wedges. It is now late June and it is about time to talk about 
% the Ice-Cream Lemma. As we shall see later, this perhaps seemingly 
% unrelated and independently interesting ice-cream lemma
% is just all we need to prove Theorem~\ref{theorem:main}.

Our main tool in proving Theorem~\ref{theorem:main} is a basic geometric lemma
that we call the ``Ice-Cream Lemma''.
Suppose that we put one scoop of ice-cream in a very large 2-dimensional cone,
such that the ice-cream touches each side of the cone at a single point.
The distances from these points to the apex of the cone are not necessarily equal.
However, we show that there is always a way of putting the ice-cream in the cone
such that they are equal.
More formally, we prove:

\begin{lemma}[Ice-cream Lemma]\label{lemma:ice}
Let $S$ be a compact convex set in the plane and fix $0 < \alpha < \pi$.
There exist a point $O$ in the plane and
two rays, $q$ and $r$, emanating from $O$ and touching $S$ 
at two \emph{single points} $X$ and $Y$, respectively, 
that satisfy $|OX|=|OY|$ and the angle bounded by $p$ and $q$ is
$\alpha$.
\end{lemma}

\begin{figure}
    \centering
    \includegraphics[width=5cm]{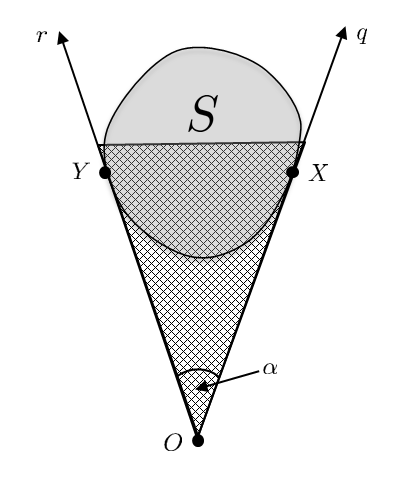}
	\caption{An illustration of the Ice-cream Lemma }
	\label{fig:ice-cream}
\end{figure}

See Figure~\ref{fig:ice-cream} for an illustration.
The requirement in Lemma~\ref{lemma:ice} that the two rays touch $S$ at \emph{single} 
points will be crucial in our proof of Theorem~\ref{theorem:main}.

At a first glance the statement in Lemma~\ref{lemma:ice} probably looks very intuitive
and it may seem that it should follow directly from a simple mean-value-theorem argument. However, this is not quite the case. Although the proof
(we will give two different proofs) indeed uses a continuity argument it is not the most
trivial one. The reader is encouraged to spend few minutes trying to come up with
a simple argument just to get the feeling of Lemma~\ref{lemma:ice} 
before continuing further.

Because the proof of Lemma~\ref{lemma:ice} is completely independent of 
Theorem~\ref{theorem:main} and of its proof, we first show, in the next section, 
how to prove Theorem~\ref{theorem:main} using Lemma \ref{lemma:ice}. We postpone
the two proofs of the Ice-Cream Lemma to the last section.

\section{Proof of Theorem~\ref{theorem:main} using Lemma~\ref{lemma:ice}}

Let $P$ be a set of $n$ points in general position in the plane.
We denote the convex hull of $P$ by $\CH{P}$,
and recall that it can be computed in $O(n\log h)$ time~\cite{KS86}.
Call two vertices of $\CH{P}$ a \emph{good pair} if there are a point $O$ and two rays $q,r$ emanating from it
creating an angle of $\frac{\pi}{3}$ such that $X \in q, Y \in r$, $|OX|=|OY|$, and $\CH{P} \subset \angle XOY$.
Note that given $\CH{P}$ and two vertices of it $X,Y$,
we can check in constant time whether $X,Y$ is a good pair.
Indeed there are exactly two points that form an equilateral triangle with $X,Y$,
and for each of these two possible locations of $O$,
we only need to check whether the neighbors of $X$ and $Y$ in $\CH{P}$ lie in  $\angle XOY$.

Lemma~\ref{lemma:ice} guarantees that $\CH{P}$ has a good pair.
Next we describe an efficient way to find such a good pair.
Suppose that $X,Y$ is a good pair.
Observe that there are two other rays that form an angle of measure $\pi/3$, contain $\CH{P}$
in their wedge, and such that one of them contains an edge of $\CH{P}$ that is adjacent
to $X$ or $Y$ and the other ray contains the other point.
(These rays can be obtained by continuously rotating the ray through $X$
while forcing the ray through $Y$ to form a $\pi/3$ angle with it, until an edge of $\CH{P}$ is hit.
Note that the distances from $X$ and $Y$ to the apex of the new wedge are no longer equal.)
Therefore, to find a good pair $X,Y$, it is enough to find for every edge $(X,X')$ of $\CH{P}$
the (at most four) points $Y$ such that $Y$ is a vertex of $\CH{P}$ and a line through $Y$
that forms $\pi/3$ angle with the line through $(X,X')$ supports $\CH{P}$.  
For every such point, we can check in constant time whether $X,Y$ or $X',Y$ is a good pair.
Finding the points $Y$ for an edge $(X,X')$ as above can be done in $O(\log h)$-time
by a binary search on the vertices of $\CH{P}$.
Thus, a good pair $(X,Y)$ and the corresponding $r$, $q$, and $O$
can be found in $O(n \log h)$-time.
% We start by applying Lemma \ref{lemma:ice} to the set $S$ taken as the 
% convex hull of $P$ and $\alpha=\frac{\pi}{3}$.
% We therefore find a point $O$ and two rays $q,r$ emanating from it
% creating an angle of $\frac{\pi}{3}$ such that there are two points $X,Y \in P$
% with $X \in q, Y \in r$, and $|OX|=|OY|$. Moreover $P \subset \wdg{q,r}$.

\begin{figure}
    \centering
    \subfigure[]{\label{fig:case1}
    {\includegraphics[width=6cm]{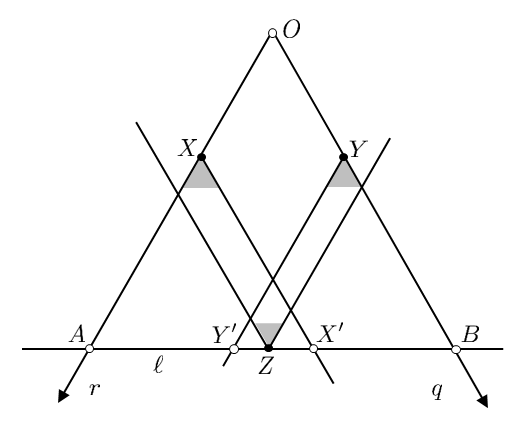}}}
        \hspace{5mm}
    \subfigure[]{\label{fig:case2}
    {\includegraphics[width=6cm]{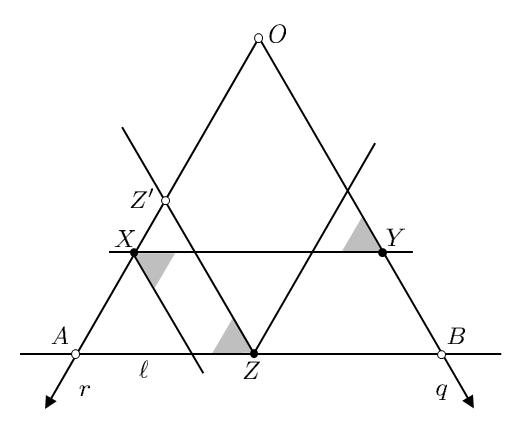}}}
	\caption{Proof of Theorem \ref{theorem:main}.}
	\label{fig:wedges}
\end{figure}

Let $\ell$ be a line creating an angle of $\frac{\pi}{3}$ with both $q$ and $r$
such that $P$ is contained in the region bounded by $q,r,$ and $\ell$ and there
is a point $Z\in P$ on $\ell$. 
(Note that $Z$ can be found in $O(\log h)$-time.)
Let $A,B$ denote the intersection points of 
$\ell$ with $q$ and $r$, respectively (see Figure~\ref{fig:wedges}).
Let $X' \in \ell$ be such that $\Delta AXX'$ is equilateral. 
Let $Y' \in \ell$ be such that
$\Delta BYY'$ is equilateral.

\noindent {\bf Case 1:} $Z \in \angle AXX'$ and $Z \in \angle BYY'$.
In this case $\measuredangle XZY \leq \frac{\pi}{3}$.
Let $W_{Z}$ be a wedge of angle $\frac{\pi}{3}$ with apex $Z$ containing
both $X$ and $Y$. Let $W_{X}$ be the wedge $\angle AXX'$ and
let $W_{Y}$ be the wedge $\angle BYY'$. See Figure~\ref{fig:case1}.
Observe that the wedge-graph
that corresponds to $W_{X},W_{Y},W_{Z}$ is connected 
($Z$ is connected by edges to 
both $X$ and $Y$). %Observe also that $W_{X} \cup W_{Y} \cup W_{Z}$
%contains $\Delta OAB$ and hence the entire set of points $P$.
%We can now easily find for each point $D \in P \setminus \{X,Y,Z\}$
%a wedge of angle $\frac{\pi}{3}$ and apex $D$ such that in the wedge-graph
%that corresponds to the set of all these edges, 
%each such $D$ will be connected to one of $X,Y,Z$.

\noindent {\bf Case 2:} Without loss of generality 
$Z \notin \angle AXX'$. In this case let $W_{Y}=\angle OYX$, 
let $W_{X}=\angle YXX'$, and let $W_{Z}=\angle AZZ'$, where $Z' \in OA$ is 
such that $\Delta AZZ'$ is equilateral. See Figure~\ref{fig:case2}.
Again we have that $X$ is connected by edges to both $Y$ and $Z$ in the 
wedge-graph that corresponds to $W_{X},W_{Y}, W_{Z}$. %We also have
%$P \subset W_{X} \cup W_{Y} \cup W_{Z}$ and we are done as in the previous 
%case.

Finally, observe that in both cases the wedge-graph contains a 2-path on the vertices $X,Y,Z$,
and $W_{X} \cup W_{Y} \cup W_{Z}$ contains $\Delta OAB$ and hence the entire set of points $P$.
We can now easily find for each point $D \in P \setminus \{X,Y,Z\}$ in constant time
a wedge of angle $\frac{\pi}{3}$ and apex $D$ such that in the wedge-graph
that corresponds to the set of all these edges, 
each such $D$ will be connected to one of $X,Y,Z$.
\bbox

It is left to prove our main tool, Lemma~\ref{lemma:ice}.
This is done in the next section.

\section{Two proofs of the ``Ice-cream Lemma''}

We will give two different proofs for Lemma \ref{lemma:ice}.

\noindent {\bf Proof I.}
In this proof we will assume that the set $S$ is \emph{strictly} convex,
that is, we assume that the boundary of $S$ does not contain a straight line
segment. We assume this 
in order to simplify the proof, however this assumption is not critical and
can be avoided.
We bring this proof mainly for its independent interest (see Claim~\ref{claim:per}
below). The second proof of Lemma \ref{lemma:ice} is shorter and applies for
general $S$.

\begin{figure}
    \centering
    \subfigure[]{\label{fig:proof1}
    {\includegraphics[width=7cm]{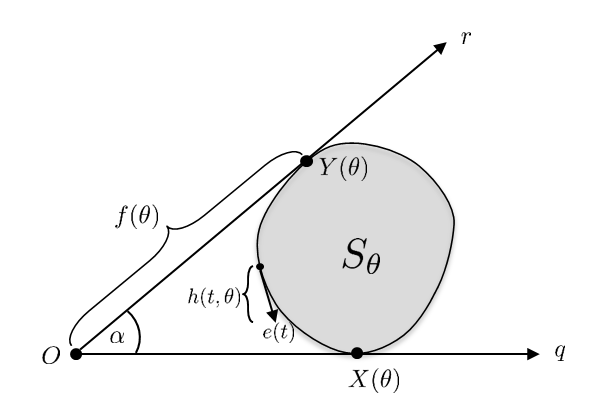}}}
        \hspace{5mm}
    \subfigure[]{\label{fig:proof2}
    {\includegraphics[width=7cm]{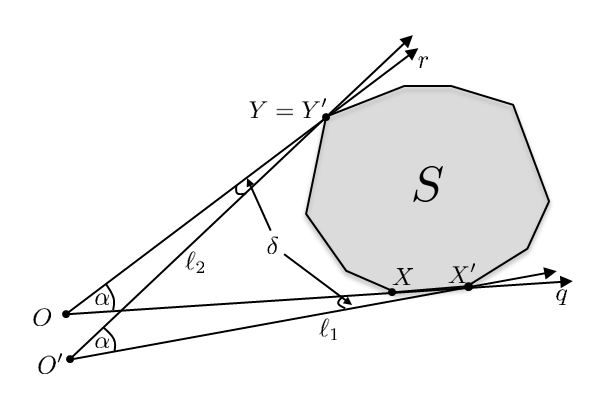}}}
	\caption{Illustrations for the proofs of the Ice-cream Lemma}
	\label{fig:ice-cream-lemma}
\end{figure}

For an angle $0 \leq \theta \leq 2\pi$
we denote by $S_{\theta}$ a (possibly translated) copy of $S$ rotated in an angle of 
$\theta$. Let $q$ and $r$ be two rays emanating from the origin $O$ and 
creating
an angle $\alpha$.
Observe that for every $\theta$ 
there exists a unique translation of $S_{\theta}$ that
is contained in $\wdg{q,r}$ and both $q$ and $r$ touch 
$S_{\theta}$ at a \emph{single} point (here we use the fact that $S$ is 
\emph{strictly} convex).
For every $0 \leq \theta \leq 2\pi$ we denote by $X(\theta)$ and $Y(\theta)$
the two touching points on $q$ and on $r$, respectively.
Let $f(\theta)$ denote the distance between $Y(\theta)$ and the
origin (see Figure~\ref{fig:proof1}). 

\begin{claim}\label{claim:per}  
$\int_{0}^{2\pi}f\,\mathrm{d}\theta=\P(S)\frac{1+\cos \alpha}{\sin \alpha}$, where $\P(S)$ denotes
the perimeter of the set $S$.
\end{claim}

\noindent {\bf Proof.}
Without loss of generality assume that the ray $q$ coincides with the 
positive part of the $x$-axis and $r$ lies in the upper half-plane. 

Consider the boundary $\partial S$ with the positive (counterclockwise) 
orientation. Note that since $S$ is convex, at each point $p\in\partial S$ there is a unique supporting line pointing forward (here after positive tangent), and a unique supporting line pointing backwards 
(the two tangent lines coincide iff 
$\partial S$ is smooth at $p$). Let $e:[0,\P(S))\to\partial S$ be a unit speed curve traveling around $\partial S$.
We define a function $h(t,\theta)$ in the following way:
If $e(t)$ belongs to the part of the boundary of $S_{\theta}$ between $X(\theta)$
and $Y(\theta)$ that is visible from $O$, then we set $h(t,\theta)$
to be equal to the length of the orthogonal projection of the unit positive tangent 
at $e(t)$ on the $y$-axis (see Figure~\ref{fig:proof1}). 
Otherwise we set $h(t,\theta)=0$.

The simple but important observation here is 
that for every $0 \leq \theta \leq 2\pi$ the expression
$\int_0^{\P(S)} h(t,\theta)\,\mathrm{d}t$ is equal to the $y$-coordinate 
of $Y(\theta)$. This, in turn, is equal by definition to 
$f(\theta)\sin \alpha$. To see this observation take a small portion of the
boundary of $S(\theta)$ of length $\mathrm{d}t$ that is visible from $O$. 
Its orthogonal projection on the $y$-axis has (by definition) length 
$h(t,\theta)\mathrm{d}t$. 
Notice that the orthogonal projection of the entire part of the boundary of $S(\theta)$
that is visible from $O$ on the $y$-axis (whose length, therefore, equals to this 
integral) is precisely all the points on the $y$-axis with smaller $y$-coordinate
than that of $Y(\theta)$.  

By Fubini's theorem we have:

\begin{equation}\label{eq:1}
\int_{0}^{2\pi}f(\theta) \sin \alpha \,\mathrm{d}\theta=
\int_{0}^{2\pi}\int_0^{\P(S)}h(t,\theta)\,\mathrm{d}t \,\mathrm{d}\theta=
\int_o^{\P(S)}\int_{0}^{2\pi}h(t,\theta)\,\mathrm{d}\theta \,\mathrm{d}t
\end{equation}
 
Moreover, for every $t$ we have:

\begin{equation}\label{eq:2}
\int_{0}^{2\pi}h(t,\theta)\,\mathrm{d}\theta=
\int_{\pi+\alpha}^{2\pi}|\sin(\theta)|\,\mathrm{d}\theta=(1+\cos \alpha).
\end{equation}

To see this observe that $e(t)$ is visible from $O$  through the 
rotation of $S$ precisely 
from where it lies on the ray $r$ until it lies on the ray $q$.
Through this period the angle which the positive tangent at $e(t)$ creates with the 
$x$-axis varies
from $\pi+\alpha$ to $2\pi$.% (see Figure~\ref{fig:?????}). 

Combining (\ref{eq:1}) and (\ref{eq:2}) we conclude:

\begin{eqnarray}
\int_{0}^{2\pi}f(\theta) \sin \alpha \,\mathrm{d}\theta & =&
\int_0^{\P(S)}\int_{0}^{2\pi}h(e,\theta)\,\mathrm{d}\theta \,\mathrm{d}t=\nonumber\\
&=& \int_0^{\P(S)}(1+\cos \alpha)\,\mathrm{d}t=\P(S)(1+\cos \alpha),\nonumber
\label{eq:4}
\end{eqnarray}

which in turn implies the desired result:
$\int_{0}^{2\pi}f(\theta)\,\mathrm{d}\theta=\P(S)\frac{1+\cos \alpha}{\sin \alpha}$.

\bbox

Analogously to $f(\theta)$ we define $g(\theta)$ to be the distance 
from $X(\theta)$ to the origin $O$. Lemma~\ref{lemma:ice} is equivalent to
saying that there is a $\theta$ for which $f(\theta)=g(\theta)$.
By a similar argument or by applying the result of Claim \ref{claim:per} to
a reflection of $S$, we deduce
that $\int_{0}^{2\pi}g(\theta)\,\mathrm{d}\theta=\P(S)\frac{1+\cos \alpha}{\sin \alpha}$.
In particular $\int_{0}^{2\pi}f(\theta)\,\mathrm{d}\theta=
\int_{0}^{2\pi}g(\theta)\,\mathrm{d}\theta$. Because  
$f$ and $g$ are continuous we may now conclude the following:

\begin{corollary}\label{corollary:equal}
Assume that $S$ is strictly convex, then 
there exists $\theta$ between $0$ and $2\pi$ such that $f(\theta)=g(\theta)$.
\end{corollary}

This completes the proof of Lemma \ref{lemma:ice} in the case where $S$ is strictly 
convex.
$\bbox$

\bigskip

We now bring the second proof of Lemma~\ref{lemma:ice}. This proof is shorter than 
the first one and does not rely on the strict convexity assumption.

\noindent {\bf Proof II.}
Consider the point $O$ such that the two tangents of $S$ through $O$ create an angle of $\alpha$ and 
such that the area of the convex hull of $\{O\} \cup S$ is maximum. By a simple compactness argument such $O$ exists.

We will show that the point $O$ satisfies the requirements of the lemma.
Let $q$ and $r$ be the two rays emanating from $O$ and tangent to $S$.
Let $X$ and $X'$ be the end points of the (possibly degenerate) line segment
$q \cap S$ and assume  
$|OX| \leq |OX'|$. Similarly, let $Y$ and $Y'$ be the two (possible equal) 
points such that
the intersection of 
$r$ and $S$ is the line segment connecting $Y$ and $Y'$ and assume 
$|OY| \leq |OY'|$.

We claim that $|OX'| \leq |OY|$ (and similarly $|OY'| \leq |OX|$).
This will imply immediately the desired result because in this case
$|OX| \leq |OX'| \leq |OY| \leq |OY'| \leq |OX|$ from which we conclude
that $X=X'$, $Y=Y'$, and $|OX|=|OY|$.

%\begin{figure}[ht]
%\begin{center}
%\input{ice_cream.pstex_t}
%\caption{Proof of Lemma \ref{lemma:ice}.}
%\label{fig:1}
%\end{center}
%\end{figure}

Assume to the contrary that $|OX'| > |OY|$. Without loss of generality assume 
that $S$ lies to the left of $q$ and to the right of $r$ 
(see Figure~\ref{fig:proof2}). Let $\delta>0$ be very small positive number and 
let $\ell_1$ be the directed 
line supporting $S$ having $S$ to its left that is 
obtained from $\overline{OX'}$
by rotating it counterclockwise at angle $\delta$. Let $\ell_{2}$ 
be the directed line supporting $S$ having $S$ to its right 
that is obtained from $\overline{OY}$
by rotating it counterclockwise at angle $\delta$ (see Figure 
\ref{fig:proof2}).

Let $O'$ be the intersection point of $\ell_1$ and $\ell_2$.
Note that $\ell_{1}$ and $\ell_{2}$ create an angle of $\alpha$.
We claim that the area of the convex hull of $\{O'\} \cup S$
is greater than the area of the convex hull of $\{O\} \cup S$.
Indeed, up to lower order terms the difference between the two equals
$\frac{1}{2}(|OX'|^2-|OY|^2)\sin(\delta) >0$. This contradicts 
the choice of the point $O$.
\bbox

\paragraph{Remarks.}
Lemma~\ref{lemma:ice} clearly holds for non-convex (but compact) sets $S$ as well,
since we can apply it on the convex hull of $S$
and observe that if a line supports the convex hull and intersects it in a single point,
then this point must belong to $S$.

From both proofs of Lemma \ref{lemma:ice} it follows, and is rather intuitive as well, that one can always find 
at least 
two points $O$ that satisfy the requirements of the lemma. 
In the first proof notice that
both functions $f$ and $g$ are periodic and therefore if they have the same
integral over $[0,2\pi]$, they must agree in at least \emph{two} distinct points, 
as they are continuous. In the second proof one can choose a point $O$
that minimizes the area of the convex hull of $O$ and $S$ and obtain a 
different solution.

% \paragraph{Acknowledgment.} We thank an anonymous referee for
% suggestions that helped improving the presentation of the paper.

\bigskip
\noindent\textit{Department of Mathematics, Physics, and Computer Science, 
University of Haifa at Oranim, Tivon 36006, Israel.
ackerman@sci.haifa.ac.il}

\bigskip
\noindent\textit{Mathematics Department,
Hebrew University of Jerusalem,
Jerusalem, Israel.
gelander@math.huji.ac.il}

\bigskip
\noindent\textit{Mathematics Department,
Technion---Israel Institute of Technology,
Haifa 32000, Israel.
room@math.technion.ac.il}

\end{document}